# Molecular mechanisms that regulate the coupled period of the mammalian circadian clock

Jae Kyoung Kim[1,*], Zachary P. Kilpatrick[2], Matthew R. Bennett[3,4], Krešimir Josić[2,5,**]

1. Mathematical Biosciences Institute, The Ohio State University, Columbus, OH 43210, USA.
2. Department of Mathematics, University of Houston, Houston, TX 77204-3008.
3. Department of Biochemistry & Cell Biology, Rice University, Houston TX, 77005.
4. Institute of Biosciences and Bioengineering, Rice University, Houston, TX, 77005.
5. Department of Biology and Biochemistry, University of Houston, Houston, TX 77204-5001.

*Email: kim.5052@mbi.osu.edu

**Email: josic@houston.edu

# Abstract

In mammals, most cells in the brain and peripheral tissues generate circadian (~24hr) rhythms autonomously. These self-sustained rhythms are coordinated and entrained by a master circadian clock in the suprachiasmatic nucleus (SCN). Within the SCN, the individual rhythms of each neuron are synchronized through intercellular signaling. One important feature of SCN is that the synchronized period is close to the population mean of cells' intrinsic periods. In this way, the synchronized period of the SCN stays close to the periods of cells in peripheral tissues. This is important because the SCN must entrain cells throughout the body. However, the mechanism that drives the period of the coupled SCN cells to the population mean is not known. We use mathematical modeling and analysis to show that the mechanism of transcription repression in the intracellular feedback loop plays a pivotal role in regulating the coupled period. Specifically, we use phase response curve analysis to show that the coupled period within the SCN stays near the population mean if transcriptional repression occurs via protein sequestration. In contrast, the coupled period is far from the mean if repression occurs through highly nonlinear Hill-type regulation (e.g. oligomer- or phosphorylation-based repression), as widely assumed in previous mathematical models. Furthermore, we find that the timescale of intercellular coupling needs to be fast compared to that of intracellular feedback to maintain the mean period. These findings reveal the important relationship between the intracellular transcriptional feedback loop and intercellular coupling. This relationship explains why transcriptional repression appears to occur via protein sequestration in multicellular organisms, mammals and *Drosophila*, in contrast with the phosphorylation-based repression in unicellular organisms. That is, transition to protein sequestration is essential for synchronizing multiple cells with a period close to the population mean (~24hr).

# Introduction

Physiological and metabolic processes like sleep, blood pressure, and hormone secretion exhibit circadian (~24hr) rhythms in mammals (1). These rhythms are mainly regulated by the master circadian clock in the suprachiasmatic nucleus (SCN) of the hypothalamus (2). The SCN consists of ~20,000 neurons, each of which exhibits rhythmic gene expression. These rhythms are mediated by an intracellular transcriptional feedback loop, in which PER/CRY dimers inhibit their own transcriptional activators, BMAL1/CLOCK dimers (3, 4). The neuronal population's rhythm is synchronized through intercellular coupling via various neurotransmitters, such as VIP, AVP, GRP, and GABA (5). In particular, experimental evidence points to VIP as a major coupling signal (6), without which SCN fails to synchronize individual rhythms (7).

Intercellular coupling within the SCN plays a pivotal role in generating robust and coherent rhythms. Individual cells within the SCN oscillate at their own periods and phases. Intercellular coupling synchronizes these rhythms, resulting in a global rhythm (7–9). Furthermore, a broad distribution of periods of individual cells becomes narrow with the coupling, which allows precise timekeeping by SCN (Fig. 1) (7, 8, 10). Coupling can also restore the rhythms among cells that lose rhythms due to mutations (11, 12). These properties of coupling within SCN have been widely explored with mathematical models. Mathematical models of SCN have shown how VIP signaling can synchronize heterogeneous rhythms (13–16) and confirmed that coupling increases the resistance of rhythms to genetic mutation (11, 12), intrinsic noise (17) and external entrainment signal (18).

One feature of intercellular coupling within the SCN that is not shared by other coupled biological oscillators (e.g. segmentation clock) (19–21) is that the coupled period, *i.e.* the global period at which all cells synchronize, is close to the population mean period of the individual cells (Fig. 1) (7, 8, 10). This feature is important because the SCN functions as a master clock that entrains peripheral clocks (1). That is, individual cells in peripheral tissues (e.g. liver and heart) generate rhythms autonomously with periods of ~24hr but are entrained by the rhythms of SCN. The closer the period of the SCN to the periods of peripheral clocks, the more likely entrainment occurs, which generates coherent systemic rhythms in the organism (1, 22). However, it is not understood what drives the period of the coupled SCN close to the population mean. Furthermore, mathematical models based on genetic feedback loops have shown significant differences (~3-6 hrs) between the coupled period and population mean, inconsistent with experimental findings (Fig. 1) (13–15).

Previous mathematical models have typically relied on Hill functions to describe transcriptional repression in the negative feedback loop (13–15). However, in a recent theoretical study it was shown that circadian clocks behave very differently when transcriptional repression occurs via protein sequestration, in which repressor inhibits a transcriptional activator via 1:1 stoichiometric binding (Fig. 2A), rather than highly nonlinear Hill-type regulation (Fig. 2B and S1) (23). That is, a model based on protein sequestration successfully reproduced various experimental observations that have not been addressed by previous models based on Hill-type regulation, such as the importance of a 1:1 molar ratio between repressor and activator and an additional negative feedback loop via Rev-erbα/β for robust circadian timekeeping (24, 25). This indicates that the mechanism of transcriptional regulation plays a key role in determining the behaviors of circadian clocks.

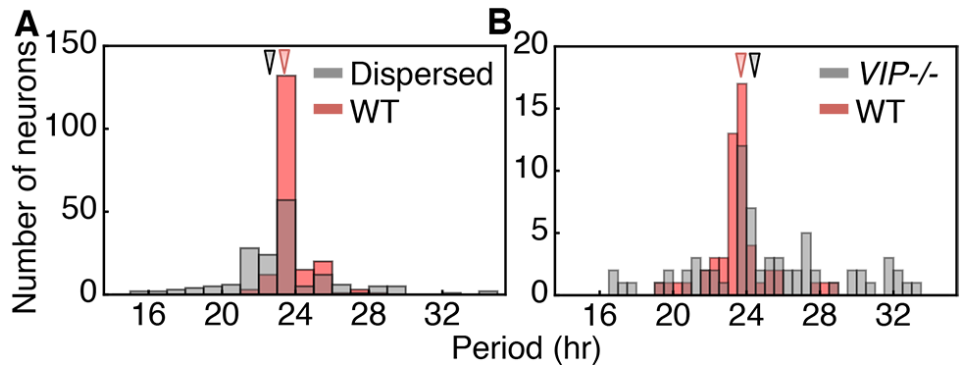

**Figure 1. Coupling maintains population mean of periods in circadian clocks.** When intercellular coupling within the SCN is disrupted either by enzymatic dispersion (**A**) or VIP$^{-/-}$ (**B**), the distributions of periods of individual neurons broadens. However the mean periods, indicated by the arrows at the top of each panel, do not significantly change when coupling is disrupted. WT: 23.3±1 and dispersed SCN: 22.7±2.9 in Fig. 1A and WT: 23.6±1.7 and VIP$^{-/-}$: 25±4 in Fig. 1B. Fig. 1A and 1B are reproduced from (8) and (7), respectively, with permission from Nature Publishing Group Ltd.

Interestingly, recent experimental studies have found that protein sequestration is responsible for repression in the negative feedback loops of circadian clocks in multicellular organisms (*Drosophila melanogaster* and mammals), which have intercellular coupling among the pacemaker cells in the brain (26–29). In contrast, a phosphorylation-based repression mechanism appears to be used in organisms which do not have this intercellular coupling. In a syncytium, *Neurospora crassa*, repressors transiently bind activators and induce phosphorylation at multiple activator sites, and thus repress its transcriptional activity (Fig. S1) (30). A similar phosphorylation-based repression mechanism is used in a unicellular organism, cyanobacteria, in which KiaA phosphorylates the multiple-sites of KiaC (31), which leads to Hill-type regulation (32, 33). These different repression mechanisms of organisms depending on the presence of intercellular coupling raises the question of whether the transition to protein sequestration is important for synchronizing the rhythms of multiple cells.

Here, we show that when transcriptional repression occurs via protein sequestration, but not Hill-type regulation, the coupled periods are near the mean period of the individual cells within the SCN. To do this, we first compare two simple mathematical models of intracellular circadian clocks, in which transcription is regulated via either protein sequestration (PS model) (23) or Hill-type repression (HT model) (34). We find that when individual oscillators are coupled in the PS model, the period of the synchronized rhythm is close to the population mean. However, the coupled period is far from the population mean in the HT model. We find that this difference is due to the functional form of the transcriptional regulation in the models – piecewise linear in the PS model and sigmoidal in the HT model. As a result, the PS model has an instantaneous phase response curve (iPRC) with balanced advance and delay regions, which leads the coupled periods to be close to the population mean. In contrast, sigmoidal transcriptional regulation leads to an unbalanced iPRC in the HT model. Finally, we find that the timescale of intercellular coupling also plays an important role in determining the coupled period. That is, the timescale of intercellular coupling needs to be faster than the timescale of the intracellular feedback loop (~24hr) to synchronize rhythms with the period close to the population mean.

We found that the mechanisms underlying the intracellular feedback loop play a pivotal role in regulating the coupled period, which reveals that two of the major functions of the SCN – the generation of a rhythm within a cell and synchronization of the rhythm across the population - are closely related. Furthermore, these findings indicate that the type of intracellular feedback mechanisms of multicellular organisms, which is different from that of unicellular organisms are necessary to synchronize rhythms of multiple cells with the population mean period (~24hr).

# Results

**Two types of gene regulation used in previous mathematical models of the circadian clocks**

To explore the role of transcriptional repression mechanisms in regulating the coupled period, we compare two main types of gene regulation, which have been used in previous mathematical models of mammalian circadian clocks: protein sequestration and Hill-type repression (Fig. 2A and B). In protein sequestration, repressors (PER-CRY) tightly bind the transcriptional activators (BMAL1-CLOCK) in a 1:1 stoichiometric complex, preventing activators from up-regulating transcription (23) (Fig. 2A). Because repressor binding inhibits the activators, the transcription rate of the repressor is proportional to the fraction of free activators. Assuming rapid binding between repressors and activators, the fraction of free activators can be described by (23, 35):

$$f(R) = \frac{A - R - K_d + \sqrt{(A - R - K_d)^2 + 4AK_d}}{2A} \xrightarrow{K_d \to 0} Max[1 - \frac{R}{A}, 0], \quad (1)$$

where $R$ and $A$ are the concentrations of repressor and activator, respectively. If the repressors bind tightly to activators ($K_d \ll 1$), the fraction of free activators can be approximated with a piecewise linear function of the molar ratio between repressors and activators (Fig. 2C). When the concentration of repressor exceeds that of the activator, the repressors strongly bind all the activators, and transcription is completely inhibited (Fig. 2C). As the concentration of repressors decreases, the activators are released proportionally to the decrease of repressors, and the transcription rate increases linearly (Fig. 2C).

Hill-type repression is another mechanism that has been used widely in models of circadian clocks since the development of Goodwin oscillator (13–16, 36, 37). In this type of gene regulation, repressors cooperatively bind to operator sites of their own gene's promoter (Fig. 2B). Assuming rapid cooperative binding, the fraction of unbound promoters as a function of repressor can be described by the Hill function (33):

$$f(R) = \frac{1}{1 + (R/K_d)^n}. \quad (2)$$

Here the relationship between the concentration of repressor and transcription rate is sigmoidal – as the repressor decreases, the transcriptional activity increases to a maximum and then saturates (Fig. 2D). Furthermore, this Hill function also describes phosphorylation-based repression mechanisms, in which the repressor phosphorylates the multiple sites of activator in a cooperative manner (Fig. S1) (33).

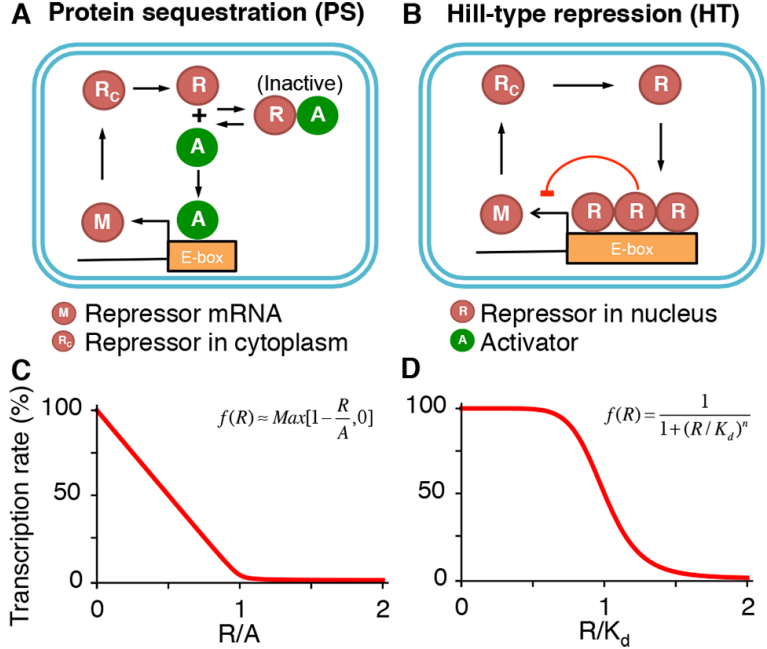

**Figure 2**. **Two types of gene regulation used in mathematical models of circadian clocks.** (**A**) Protein sequestration has been used to model the repressor (Per1/2 and Cry1/2) gene regulation, in which the repressors inactivate the activators (BMAL1-CLOCK) via 1:1 stoichiometric binding. (**B**) Hill-type regulation of repressor transcription is derived assuming fast cooperative binding reactions, such as oligomeric binding of repressor proteins to their own promoter. Phosphorylation-based repression can also induce Hill-type regulation (Fig. S1). (**C**) Transcriptional regulation via protein sequestration has an approximately piecewise linear relationship between repressor and transcription rate. (**D**) Hill-type regulation yields a sigmoidal relationship.

**Mathematical models of the intracellular feedback loop in a single cell**

The two different types of gene regulation lead to two different mathematical models of the transcriptional negative feedback loop in circadian clocks (Fig. 2A and B) (23, 34, 38),

$$\frac{dM}{dt} = \alpha_1 f(R) - \beta_1 M,$$
$$\frac{dR_c}{dt} = \alpha_2 M - \beta_2 R_c,$$
$$\frac{dR}{dt} = \alpha_3 R_c - \beta_3 R. \quad (3)$$

Here $M$, $R_c$, and $R$ represent the concentration of repressor mRNA, cytoplasmic repressor protein, and nuclear repressor protein, respectively. In Eq. (3) mRNA is first translated into repressor protein in the cytoplasm. This protein then enters the nucleus and inhibits its own transcription either through protein sequestration (PS model) or Hill-type regulation (HT model). Thus, the difference between the two models is the form of the mRNA transcription rate, $f(R)$ (Eq. (1) or (2)). We next put the models into dimensionless form, reducing the number of parameters: Scaling of $M$, $R_c$, and $R$ normalizes all production rates ($\alpha_i$). Furthermore, we assume that the clearance rates of all species ($\beta_i$) are equal, increasing the chance of oscillations in the Goodwin oscillator (39, 40). With this assumption and non-dimensionalization of time, all the clearance rates ($\beta_i$) can be normalized. The resulting models have two free parameters, both in the transcription function, $f(R)$ (23):

$$\frac{dM}{dt} = f(R) - M,$$

$$\frac{dR_c}{dt} = M - R_c,$$

$$\frac{dR}{dt} = R_c - R. \quad (4)$$

In the PS model, the two parameters, the activator concentration ($A$) and dissociation constant between the activator and repressor ($K_d$) in the Eq. (1), determine the dynamics of model. In the HT model, the Hill-coefficient ($n$) and dissociation constant between the repressor and gene promoter ($K_d$) in Eq. (2) govern the dynamics of HT model. The parameter values for the PS model are selected to be the same as in the original model (23). The two free parameters in the HT model are selected such that the oscillatory solutions in both models have the similar periods and amplitudes (Fig. S2).

**Mathematical models of intercellular coupling among multiple cells**

Next, we couple multiple single cell models through the intercellular signal, VIP ($V$) (Figure 3).

$$\frac{dM_i}{dt} = f(R_i) - M_i + \frac{\mu}{N} \sum V_i,$$

$$\frac{dR_{ci}}{dt} = M_i - R_{ci},$$

$$\frac{dR_i}{dt} = R_{ci} - R_i,$$

$$\frac{dV_i}{dt} = \tau(f(R_i) - V_i), \quad (5)$$

where $N$ denotes the number of cells, each indexed by $i=1,…,N$. In this model, each cell releases VIP into the extracellular space at a rate proportional to the activity of the promoter, $f(R)$ (41, 42). Similar to previous models, we assume that VIP in the extracellular space enters each cell at an equal rate because the VIP diffusion is fast relative to the period (~24hr) (13, 43). Once it has entered the cell, VIP promotes the transcription of the repressor gene (44–46). The parameters $\mu$ and $\tau$ describe the coupling strength and the timescale of intercellular coupling, respectively.

When $\mu=1$, the strength of coupling-induced transcription of the repressor ($\frac{\mu}{N}\sum V_i$) is similar to that of the intracellular feedback loop ($f(R)$). Experimental data indicates $\mu<1$ because transcription of *Per2* is increased by ~20-40% upon VIP treatment (47). The timescale of intercellular coupling, $\tau$, represents how quickly VIP is released from cells and activates transcription in neighboring cells relative to the intracellular feedback loop (~24hr). Previous experiments showed the release of VIP peaks within 30 min after the SCN is subjected to a light pulse or treated with the glutamate agonist, *N*-methyl-D-aspartate (NMDA) (48). Furthermore, approximately 5 min after VIP treatment, active CREB reaches its peak in rat anterior pituitary (49). Overall, these experimental data indicate that intercellular coupling occurs much faster, compared to the intracellular feedback loop (~24hr), so we use $\tau=20$ to describe the fast coupling process. With this choice of coupling timescale, the production rate ($f(R)$) and the degradation rate ($V_i$) of coupling signal have a similar phase in the model. Thus, the rhythm of mRNA produced by intracellular feedback loop, $f(R)$ and the rhythm of mRNA produced by intercellular coupling, $\frac{\mu}{N}\sum V_i$ show similar phases in the model. This matches recent experimental findings showing that the peaks of *Per1* transcription through the intracellular feedback loop and intercellular feedback loop are close to each other (46).

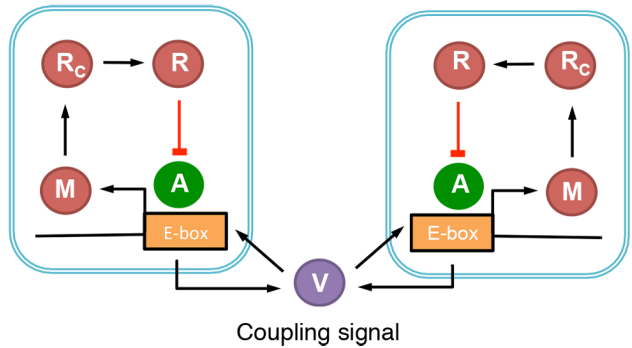

**Figure 3**. **Description of intercellular coupling in the model.** The coupled signal (VIP) is rhythmically released into the extracellular space and then enters all cells at equal rate and promotes the transcription of repressor through the CREB promoter.

**The coupled periods of a fast cell and a slow cell**

The main question we address in this work is how the periods of individual cells change in the presence of coupling. To start, we consider a pair of cells with different intrinsic periods. Following previous studies we scale time differently in two copies of a model cell (Eq. (5)) to achieve a difference in periods (13, 16). All production and degradation rates are divided by a rescaling factor, 1 and 1.2, respectively, resulting in periods that differed by 20%. After coupling the two model cells, we estimated the change in period using a Fast Fourier Transform (FFT) (Fig. 4A and B). As the coupling strength increases, the rhythms are synchronized in both the PS and HT models. However, the two models differ in how the frequency of each oscillator changes after coupling. As the coupling strength increases in the PS model, the frequencies of both the fast and slow cells tend toward the mean frequency of the two uncoupled cells (Fig. 4A). Once cells synchronize, the coupled frequency is close to the mean in the PS model. In contrast, in the HT model, the frequency of the slow cell tends to increase much more than the frequency of the fast cell decreases (Fig. 4B). This trend continues until the two cells synchronize at a frequency significantly above the mean. This difference between the PS and HT model does not depend on parameter choice (Eq. (5)) (Fig. S3). Furthermore, when we use Michaelis-Menten type coupling (13, 14) rather than linear coupling of VIP between cells in Eq. (5), we obtain similar results (Fig. S4). Next, we test whether the behavior of the HT model is due to the high Hill-coefficient. We extended the HT model to include more intermediate reaction steps, which allows the model to oscillate with a lower Hill coefficient (39). Even with a low Hill-coefficient the coupled period of the model is still far from the mean period in this extended HT model (Fig. S5).

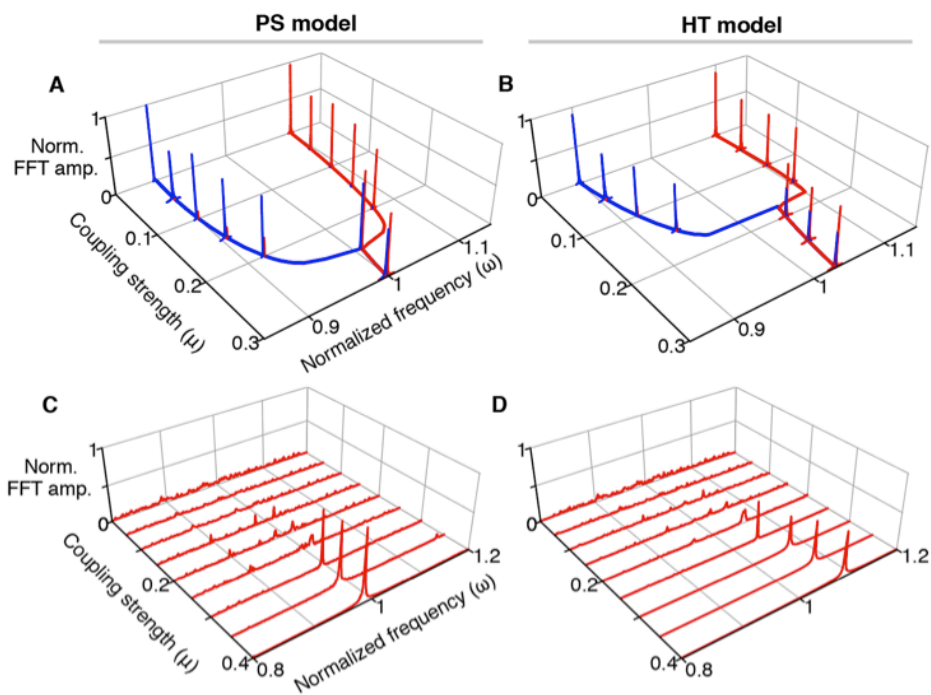

**Figure 4. The coupled periods of heterogeneous cells in the PS and HT model.** (**A**) When a fast cell and a slow cell are coupled, the coupled frequency, at which two cells synchronize, of the PS model are similar to the mean frequency of uncoupled cells, but (**B**) that of HT model is greater than the mean frequency. Frequencies are estimated using a FFT and normalized to make the mean frequency unity. Two different frequencies of single cell models are obtained by dividing all production and degradation rates by common rescaling factors of 1 and 1.2, respectively. These results are robust to parameter changes (Fig. S3 and 5) and the introduction of nonlinearities in the coupling (Fig. S4). Here, we represent the results involving $\mu$=0, 0.05, 0.1, …, 0.3. (**C**) When 100 cells with different periods are coupled, the coupled frequencies of the PS model converge to the mean frequency of uncoupled cells, but (**D**) those of the HT model become larger than the mean frequency. Here, 100 rescaling factors for different frequencies are drawn randomly from a normal distribution of mean 1 and standard deviation 0.15, matching the experimental data (Fig. 1A and B). Similar to Fig. 4A and B, frequencies are normalized, so that the mean frequency becomes 1. See Fig. S6 for stronger coupling strengths and Fig. S7 for cell populations with larger heterogeneities.

**The coupled periods of heterogeneous cell population**

Next, we test how the coupling affects the periods in a population of 100 cells with different intrinsic periods. To achieve heterogeneity we again rescale the time of each cell in the population using 100 rescaling factors sampled from a normal distribution with mean 1 and standard deviation 0.15. This generates variability in periods similar to that observed in the SCN (Fig. 1) (7, 8, 50). In the PS model, increasing coupling strength again causes the frequencies of individual cells to cluster around the population mean of the uncoupled cells (Fig. 4C). When coupling strength exceeds a threshold ($\mu$~0.3), rhythms are synchronized with frequencies close to the population mean (Fig. 4C). However, in the HT model, as coupling strength increases, frequencies cluster around a value greater than the mean frequency of the uncoupled cells (Fig. 4D). When the coupling strength exceeds a threshold ($\mu$~0.3), the population synchronizes at a frequency significantly above the population mean (Fig. 4D). The frequencies of the coupled HT model only approach the population mean when the coupling strength far exceeds the threshold ($\mu$~1) (Fig. S6). With coupling this strong, the coupled frequencies of the PS model become slightly smaller than the mean frequency (Fig. S6). However, experimental evidence suggests that the coupling strength is much smaller than unity (47). Furthermore, the HT model can exhibit synchronous oscillations at a frequency close to the population mean only at unrealistically large coupling strength. This would require a large amount of neurotransmitters at a high cost to the organism. Thus, the PS model with the weak coupling provides a more efficient mechanism for synchrony than the HT model with strong coupling. We also examine systems of oscillators whose distribution of uncoupled periods had a larger variability. Even in this case, the PS model synchronizes at frequencies that are close to the population mean (Fig. S7A). However, the frequencies of the synchronized HT model are again much larger than the mean with the realistic coupling strength (Fig. S7B).

**iPRCs and AIFs of PS model and HT model**

We have shown that the period of the synchronous population is close to the population mean for the PS model, but significantly shorter than the population mean for the HT model (Fig. 4). To understand the mechanisms that underlie this difference, we employ the theory for weakly coupled oscillators, which has been used widely to understand synchronization in networks of oscillators (51–53). Assuming weak coupling, the theory allows us to describe the essential dynamics of the four-dimensional cell model (Eq. (5)) using a single differential equation for the phase of the limit cycle. To derive the equation for the phase dynamics, first, we need to estimate the iPRCs, $Z(\theta)$, in response to mRNA perturbation:

$$Z(\theta) = \lim_{\Delta M \to 0} \frac{\Delta \theta}{\Delta M},$$

where $\Delta M$ represents the brief perturbation of mRNA and $\Delta \theta$ represents the phase change due to the perturbation of mRNA. Numerically we calculate the iPRCs for both the PS and HT models (Fig. 5A and B) (54). The advance and delay region of the iPRC are balanced in the PS model (Fig. 5A), whereas the advance regions of the iPRC is much larger than the delay region in the HT model (Fig. 5B). Importantly, the iPRC of the PS model more closely resembles the experimentally measured PRC than does the iPRC of the HT model. When the PRC is measured in response to 100 nM VIP in the SCN, the delay region is slightly larger than the advance region (47).

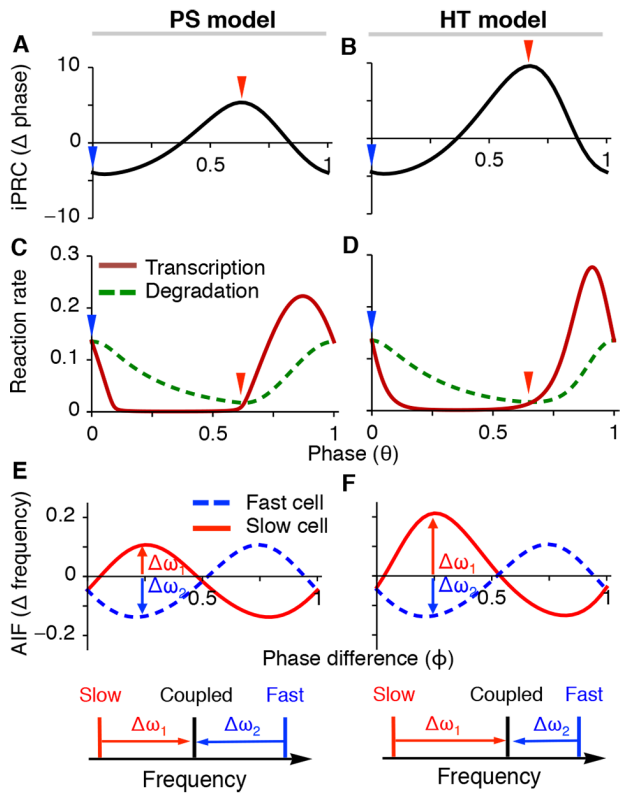

**Figure 5 iPRCs and AIFs explain the different coupled periods of the PS and HT models.** (**A**) The iPRC of the PS model has similarly sized advance and delay regions while (**B**) the iPRC of HT model has a larger advance region. (**C**) The steepness of transcription rates of PS model are similar when iPRC attains a maximum or minimum whereas (**D**) those of HT model show significant differences. (**E**) The AIF of the PS model predicts that the frequencies of the fast cell and slow cell change with a similar magnitude when they are synchronized, (**F**) but the AIF of HT model predicts that a slow cell has a larger frequency change than the fast cell, matching previous simulations (Fig. 4). See Fig. S8 for AIFs with different parameter choices.

Next, we explore why the iPRC of the PS model is more balanced than that of the HT model. To do this, we analyze the magnitudes of the maxima and minima of the iPRCs in the models, which indicate the largest phase advance and delay, respectively. We found that the extrema of the iPRC occur when the time derivative of the mRNA is zero – *i.e.* when the transcription and degradation rates of the mRNA are equal (Fig. 5C and D). This occurs because the phase of the oscillation is most sensitive to mRNA changes when the time derivative of the mRNA is zero (see the Appendix). Furthermore, we found that the slope of the transcription rate at these times appears to determine the extrema of the iPRC. That is, the maximum and minimum of the iPRC is approximately proportional to the inverse of the slope of transcription rates (see Appendix for details)

$$Z(\theta)_{extrema} \approx \frac{1}{\beta} \frac{-W(-e^{-e^{-T}-T})}{1+W(-e^{-e^{-T}-T})} \propto \frac{1}{\beta},$$

where $\beta$ is the slope of transcription rates at the phase when the iPRC is extrema, $T$ is the reference time, and $W$ is the Lambert $W$ function – a branch of the inverse of $We^W$. When the iPRC attains its maximum or minimum, the steepness of transcription rates are similar in the PS model (Fig. 5C) because the transcriptional regulation curve is piecewise linear (Fig. 2C). However, the steepness of transcription rates are very different in the HT model (Fig. 5D) because the transcriptional regulation curve is sigmoidal (Fig. 2D). In particular, when the iPRC is close to the maximum, the slope of the transcription rate is very small in the HT model, resulting in an iPRC with a large maximum in the HT model. Therefore, the PS model has a balanced iPRC due to its approximately piecewise linear gene regulation, but the HT model has an iPRC with a larger advance region due to sigmoidal gene regulation.

Finally, by convolving these iPRCs with the coupling signal, we can estimate the equations for the phases of two coupled heterogeneous cells (53).

$$\frac{d\theta}{dt} = \omega + \frac{\mu}{Period}\int_0^{Period} Z(\tilde{t})\frac{V(\tilde{t})+V'(\tilde{t}+\theta'-\theta)}{2}d\tilde{t} = \omega + \mu H(\theta'-\theta),$$
$$\frac{d\theta'}{dt} = \omega' + \frac{\mu}{Period}\int_0^{Period} Z(\tilde{t})\frac{V(\tilde{t})+V'(\tilde{t}+\theta-\theta')}{2}d\tilde{t} = \omega' + \mu H(\theta-\theta'). \quad (6)$$

Here $\omega$ and $\omega`$ represent the two different frequencies of the individual cells, and the average interaction function (AIF), $H(\theta'-\theta)$ describes the phase change due to the coupling signal. Furthermore, by subtracting these two phase equations (Eq. (6)), we can generate an equation for the phase difference $\phi = \theta'-\theta$.

$$\frac{d\phi}{dt} = w'-w + \mu(H(-\phi)-H(\phi)) = \Delta\omega + \mu(H(-\phi)-H(\phi)). \quad (7)$$

When synchronization occurs, the phase difference ($\phi$) will reach steady state and the left side of Eq. (7) will become zero, so that

$$\Delta\omega = \mu(H(\phi)-H(-\phi)). \quad (8)$$

The solution ($\phi$) of Eq. (8) describes the stable and constant phase difference of two cells when they are synchronized. Therefore, when synchronization occurs, due to these stable and constant phase differences, the AIFs become constant in Eq. (6). These constant AIFs quantify frequency change of individual cells due to coupling. The positive (negative) part of the AIF represents the speeding up (slowing down) of frequency due to coupling. The AIFs of the PS model show a balance between the positive and negative regions. Thus, a slow cell and a fast cell show similar changes in their frequencies to the coupled frequency when they are synchronized (Fig. 5E), matching previous simulations (Fig. 4A and C). Furthermore, the balanced AIFs of the PS model do not depend on our choice of parameters (Fig. S8), explaining why the coupled frequencies are close to the mean frequency regardless of parameter choice (Fig. S3). However, the AIFs in the HT model have a positive region that is larger than the negative region. Thus, coupling will alter the frequency of a slow cell more than a fast cell (Fig. 5F), matching our simulations (Fig. 4B and D).

In summary, the PS model has balanced iPRC and AIF due to piecewise linear gene regulation, but the HT model has an iPRC and AIF with a larger positive region due to sigmoidal gene regulation. These results show the tight relationship between the intracellular feedback loop (the shape of transcription regulation curves) and intercellular coupling (AIFs) (Fig. 2 and 5).

**Relationship between the timescale of the coupling signal and coupled periods**

We have assumed that the timescale of intercellular coupling is faster than the timescale of the intracellular feedback loop in the model (Eq. (5)) matching experimental data of circadian clocks (46, 48, 49). Interestingly, in somite clocks, which regulate the segmentation timing in developing embryos, the timescale of intercellular coupling is similar to that of the intracellular feedback loop (19, 55). Previous studies have found that delays in coupling can produce coupled periods that differ significantly from the mean period (19). We test whether the speed of the coupling signal affects the coupled periods in our models. Similar to somite clocks, in both the PS and HT models, the coupled frequencies become significantly smaller than the mean frequency when the time scale of the coupling signal is similar to that of the intracellular feedback loop (Fig. 6). Furthermore, with slow coupling, stronger coupling strengths are required for synchronization than with fast coupling (Fig. 4C and D). In particular, the oscillations of HT model cease when coupling strengths are between 0.4 and 0.7, a phenomenon known as oscillation death (Fig. 6B). This suggests coupling should be fast for a population of cells to synchronize at the mean frequency of the uncoupled cells.

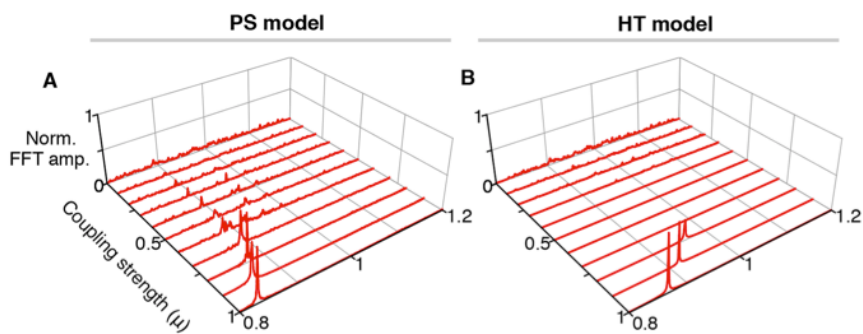

**Figure 6. When the speed of coupling signals is slow, the coupled periods become longer than mean periods.** Here, $\tau=1$. Other parameters are the same as Fig. 4C and D.

**Discussion**

The mammalian circadian clock is a hierarchical system, in which the master clock in SCN functions as a pacemaker and synchronizer of peripheral clocks to generate a coherent systemic rhythms throughout the body (5, 6). The synchronization between the master and peripheral clocks is more likely if their periods are close, which can be achieved when the coupled period of the SCN is close to its population mean period (Fig. 1). In this work, we found that such parity can be achieved if two important mechanisms of intra- and intercellular feedback loop are present in the SCN: transcriptional regulation via protein sequestration and intercellular coupling that is fast compared to the intracellular feedback loop. Using the theory of weakly coupled oscillators (51–53), we found that the piecewise linear gene regulation with protein sequestration (Fig. 2C) is key for the population to synchronize with the population mean period. When transcriptional regulation is piecewise linear (Fig. 2C) rather than sigmoidal (Fig. 2D), the iPRCs and AIFs are balanced, and the coupled periods are close to the population mean period (Fig. 6). Interestingly, only protein sequestration appears to lead piecewise linear gene regulation among other proposed rhythm generating mechanisms like oligomerization, multiple phosphorylation, and cooperative enzyme kinetics (32). However, it would be interesting to examine further if such exist and study their effect on synchronous oscillations.

Circadian clocks are widely found in organisms as diverse as bacteria, algae, plants, fungi, insects, and mammals (56). Whereas each of these organisms appear to use an intracellular negative feedback loop to generate circadian rhythms, there is a variety of mechanisms by which negative feedback is mediated. In mammals and *Drosophila*, the repressor (PER) appears to inhibit the activator (BMAL1-CLOCK in mammals and CYC-CLK in *Drosophila*) through protein sequestration. In both of these clocks, repressors tightly bind activators in a 1:1 stoichiometric complex, prohibiting activators from binding DNA (26–29). In contrast, a phosphorylation-based repression mechanism appears to be used in *Neurospora crassa* (Fig. S1). Here, the repressor (FRQ) binds the activator (WC complex) transiently and recruits kinases, which phosphorylate multiple sites of the activator (WC complex) and represses the transcriptional activity of the activator (30). Furthermore, the repressor concentration in *Neurospora* is much lower than that of the activator in nucleus, as kinase at low concentration is usually enough to phosphorylate its substrate (57–59). A similar phosphorylation-based repression mechanism is used in cyanobacteria, in which KiaA phosphorylates the multiple-sites of KiaC (31). Taken together, protein sequestration appears to be used as a repression mechanism in multicellular organisms, mammals and *Drosophila,* but not in a syncytium, *Neurospora* and a unicellular organism, cyanobacteria. This raises the question of why different mechanisms are used for transcriptional regulation depending on the type of organisms. It is known that repression through phosphorylation at multiple sites results in Hill-type regulation (32, 33) (Fig. 2D and S1).,which cannot maintain the mean frequency after intercellular coupling with an excitatory neurotransmitter (Fig. 4B and D). Our work indicates that a transition from phosphorylation-based repression to protein sequestration may be necessary to synchronize rhythms of multiple cells through the intercellular coupling at a population mean period (~24hr) (Figs. 4 and 7).

We have shown that protein sequestration is required to synchronize rhythms at the mean period (Fig. 4) under a specific type of coupling used in mammalian circadian clocks (Fig. 3). Different types of coupling may require different types of transcriptional regulation to achieve such synchrony. Indeed, a recent modeling study shows that phosphorylation-based repression can also lead to synchrony of multiple rhythms at the mean period when a common enzyme is shared for phosphorylation, which leads a intracellular coupling (60). This type of coupling appears to occur in the circadian clocks of a multinucleate system, *Neurospora*, in which cytoplasm and organelles, including nuclei, move between compartments due to an incomplete cross wall (61). Indeed, fused strains of *Neurospora* circadian clock synchronize multiple rhythms at their mean period even with phosphorylation-based repression (62). It would be interesting to explore how transcriptional regulation affects the coupled period under various coupling mechanisms.

Many coupled oscillator systems other than circadian oscillators have been identified in biological systems (63). *Dictyostelium discoideum* cellular oscillators are coupled by an intercellular cAMP signal (64). In the vertebrate embryo, Her1 rhythms are synchronized through the Delta-Notch pathway (65). $Ca^{2+}$ oscillations are also synchronized by gap junctions in pancreatic $\beta$ cells (20). In contrast to circadian clocks, the periods of these systems change greatly with coupling strength. When a major coupling signal is disrupted, the mean periods of the cell population increase by ~20% in the segmentation clocks of the zebrafish *Danio rerio* (19). Furthermore, when cell density is dispersed, which reduces the strength of coupling, the periods of cAMP oscillations in *Dictyostelium* and $Ca^{2+}$ oscillations in pancreatic $\beta$ cells change by more than 50% (20, 21). These results suggest that circadian clocks include mechanisms that tightly regulate the period of synchronized oscillations. Interestingly, the intercellular coupling appears to be faster than the timescale of the intracellular feedback loop in circadian clocks (48, 49) whereas these timescales are similar in other cellular oscillators (19, 55, 64). Further, circadian clocks (~24hr) have a much longer period than other coupled cellular oscillators (~3-20min), which makes it more plausible that the coupling process occurs much faster than the intracellular feedback loop. Together with the above experimental evidence, our simulation results (Figs. 6) indicate that fast intercellular coupling in the circadian clock plays an important role in synchronizing oscillation at the population mean period.

In the present work, we focused on the repression mechanisms of transcription, and used simple models of other process (Fig. 2A and B). We expect that our results hold in more complex systems because our mathematical analysis shows how the shape of gene regulation curves affects the PRC and coupled period regardless of the complexity of the system (Fig. 5). Indeed, more complex models of circadian clocks based on Hill-type gene regulation exhibit a large difference between coupled and mean period (13–15). It would be an interesting future work to see whether the detailed model based on protein sequestration can exhibit the coupled period similar to the mean period (23, 66). Recent studies have found that neurotransmitters other than VIP (e.g. GABA and AVP) appear to be involved in the synchronization (5, 6, 67). Furthermore, different neurotransmitters are released depending on the region of SCN (5, 6). Future models should include various neurotransmitters and spatial heterogeneity with various network architecture of coupling (15). Testing our predictions *in vivo* might be difficult, but synthetic biology could provide a reasonable alternative (68). Transcriptional regulation with the protein sequestration has been successfully generated in a synthetic system (35). Furthermore, synthetic systems capable of synchronizing genetic clocks through quorum sensing have been developed (69). It would be interesting to extend this work to test the role of protein sequestration and intercellular coupling timescale in the regulation of coupled cellular periods using synthetic biology.

**Appendix**

**Analysis of iPRC of PS model and HT model**

There are various methods for estimating the iPRCs numerically (54). However, the closed form of iPRCs have been analyzed in very limited examples, such as a simple integrate-fire model or normalized forms of periodic firing neuronal models near bifurcation points (70, 71). The dynamics of our model occurs in multiple dimensions with a non-linearity, making direct iPRC analysis difficult. To analyze the iPRC of our model, we approximate the non-linear term of our model (Eq. (5)), $f(R)$ to be a linear function of time, making our model similar to an integrate-and-fire type model

$$\frac{dM}{dt} = (\alpha + \beta t) - M(t).$$

This approximation works for an appreciable portion of the time domain in the PS model because overall transcription rate can be approximated as a piecewise-linear function of time (Fig. 5C). However, in the HT model, the transcription rate changes non-linearly as time changes (Fig. 5D), so this approximation works only locally. Furthermore, this approximation reduces the system to a single dimension, which does not consider the effect of other variables on mRNA concentration. Because the iPRC considers small perturbations of mRNA concentration, the dynamics will relax quickly to the original limit cycle. With this approximation, let us estimate the maximum and minimum of iPRC. As we discussed in the results section, the extrema of the iPRC occur when the derivative of mRNA is zero or the inverse derivative of mRNA is infinite, so that the response of time or phase to mRNA perturbation is most sensitive (Fig. 5C and D). If we set the time to be zero when the iPRC is maximum or minimum, the initial condition should be $M(0)=\alpha$, which makes the derivative of mRNA zero initially

$$\frac{dM}{dt} = (\alpha + \beta t) - M(t), \quad M(0) = \alpha. \tag{9}$$

Now, let us consider the small perturbation of mRNA, which can be represented as perturbed initial condition

$$\frac{d\bar{M}}{dt} = (\alpha + \beta t) - \bar{M}(t), \quad \bar{M}(0) = \alpha + \varepsilon. \tag{10}$$

Then, we can easily find the solutions of original system (Eq. (9)) and perturbed system (Eq. (10))

$$M(t) = \alpha - \beta + \beta t + \beta e^{-t},$$
$$\bar{M}(t) = \alpha - \beta + \beta t + \beta e^{-t} + \varepsilon e^{-t}. \quad (11)$$

If the perturbation changes phase with $\Delta\theta$, these two solutions will have following relationship at the reference time, $T$

$$M(T) = \bar{M}(T - \Delta\theta). \quad (12)$$

In the integrate-and-fire model, the firing time is considered as the reference time, $T$ (70). However, it is difficult to define $T$ clearly in our model. $T$ should be small enough for the local approximation of our model to work. Furthermore, $T$ should be long enough so the perturbed system relaxes sufficiently close to the original system. The accurate estimation of T is essential to calculate $\Delta\theta$, but not for our purpose (see below). By substituting the solutions (Eq. (11)) to the Eq. (12)), we can get

$$\alpha - \beta + \beta T + \beta e^{-T} = \alpha - \beta + \beta(T - \Delta\theta) + \beta e^{-(T-\Delta\theta)} + \varepsilon e^{-(T-\Delta\theta)},$$
$$\beta e^{-T} = -\beta\Delta\theta + (\beta + \varepsilon)e^{-(T-\Delta\theta)}.$$

This equation includes $\beta$, but not $\alpha$, indicating that $\Delta\theta$ will depend on only $\beta$, the slope of transcription rate. Because this equation includes exponential function, the solution can be expressed by using Lambert $W$ function, which is a branch of the inverse of $We^W$.

$$\Delta\theta(\varepsilon, T) = e^{-T} - W(-(1+\varepsilon/\beta)e^{-e^{-T}-T}).$$

Now let us estimate the extrema of iPRC:

$$Z(\theta)_{max/min} = \lim_{\varepsilon \to 0} \frac{\Delta\theta(\varepsilon, T)}{\varepsilon} = \lim_{\varepsilon \to 0} \frac{\partial \Delta\theta(\varepsilon, T)}{\partial \varepsilon},$$

in which the last equality comes from l'Hôpital's rule. By applying the property of Lambert $W$ function,

$$\frac{dW(z)}{dz} = \frac{W(z)}{z(1+W(z))},$$

we can get

$$\lim_{\varepsilon \to 0} \frac{\partial \Delta\theta(\varepsilon, T)}{\partial \varepsilon} = \lim_{\varepsilon \to 0} \frac{e^{-e^{-T}-T}}{\beta} \frac{1}{-(1+\varepsilon/\beta)e^{-e^{-T}-T}} \frac{W(-(1+\varepsilon/\beta)e^{-e^{-T}-T})}{1+W(-(1+\varepsilon/\beta)e^{-e^{-T}-T})}.$$

Finally, we find that the extrema of iPRC are approximately proportional to the inverse of slope of transcription rate.

$$Z(\theta)_{max/min} \approx \lim_{\varepsilon \to 0} \frac{\Delta\theta(\varepsilon,T)}{\varepsilon} = \frac{1}{\beta} \frac{-W(-e^{-e^{-T}-T})}{1+W(-e^{-e^{-T}-T})} \propto \frac{1}{\beta}.$$

**Supporting material**

Eight figures are available at Supporting Material. A reference (72) appears in the Supporting Material.

**Acknowledgement**
We thank Daniel Forger, Daniel DeWoskin, and Michael Schwemmer for valuable discussion about this manuscript; and Sato Honma and Daisuke Ono for providing quantifying data in Figure 1A. This work was funded by the NIH, through the joint NSF/NIGMS Mathematical Biology Program Grant No. R01GM104974 (MRB and KJ), NSF grants DMS-1311755 (ZPK) and DMS-1122094 (KJ), the Robert A. Welch Foundation Grant No. C-1729 (MRB) and NSF grants DMS-0931642 to the Mathematical Biosciences Institute (JKK).

# Supporting Figures

**Table of Contents**

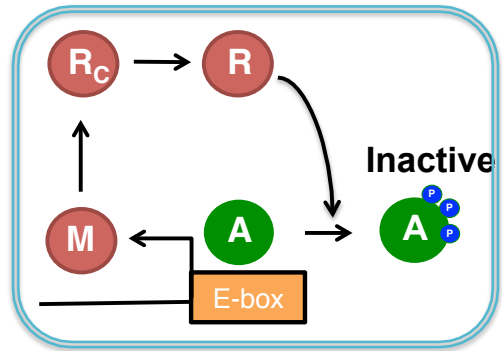

**Figure S1. Phosphorylation-based repression in *Neurospora* circadian clock.** In *Neurospora Crassa* circadian clock, repressor protein (FRQ) recruits kinases that phosphorylate multiple sites of activator protein (WCC). This phosphorylation inactivates the activator protein. Whereas this phosphorylation-based repression is different from the direct repression via oligomerization, which appears in the Goodwin oscillator (Fig. 2B), both mechanisms can be modeled using Hill equations (1). Thus, the Goodwin oscillator has been used as a *Neurospora* circadian clock model (2). The Hill-coefficient represents the number of phosphorylation sites in phosphorylation-based repression mechanism (1)**.**

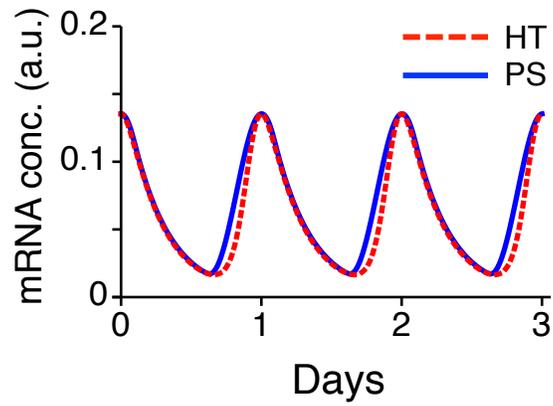

**Figure S2. mRNA timecourses of PS model and HT model.** Parameters of PS model ($K_d$ and $A$) and HT model ($K_d$ and $n$) are selected so the two models have similar amplitudes and periods. Here, $A=0.0659$ and $K_d=10^{-5}$ for PS model and $n=11$ and $K_d=4\times10^{-2}$ for HT model.

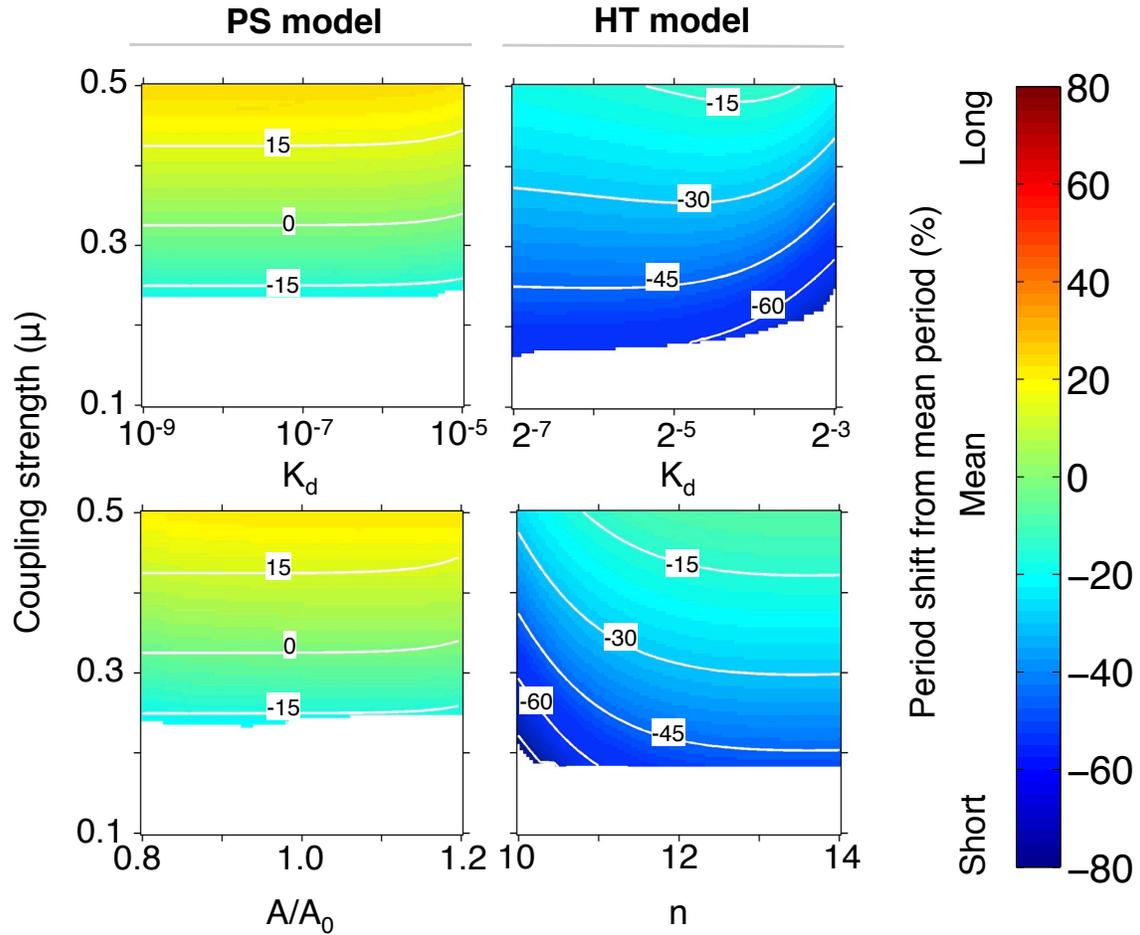

**Figure S3. The coupled periods of two heterogeneous cells with different parameters.** The shifts of coupled periods from the mean period are represented with color and contour. Regardless of the choice of parameters, the coupled periods of the PS model are similar to the mean period, but those of HT model are shorter than the mean period with a weak coupling ($0.2<\mu<0.4$). The coupled periods of HT model are similar to the mean period only when coupling becomes strong ($\mu>0.5$). Here, $A_0$ indicates the concentration of activator in the original model. Rescaling factors are 1 and 1.2 for two cells as in Fig. 4A.

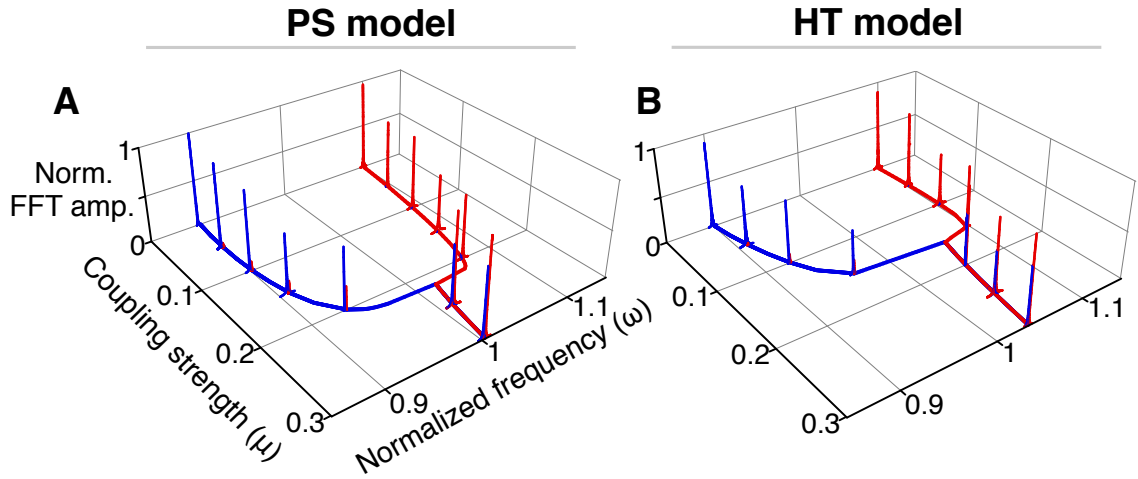

**Figure S4. Coupled periods of two heterogeneous cells via Michaelis-Menten type coupling.** When Michaelis-Menten type equations were used for coupling, results are similar to Fig. 4A-B, where linear coupling is used.

$$\mu \sum_{i=1}^{N} V_i / N \to \frac{\mu \sum_{i=1}^{N} V_i / N}{K_C + \sum_{i=1}^{N} V_i / N}$$

Here, $Kc$=0.7 and other parameters were the same with Fig. 4A-B.

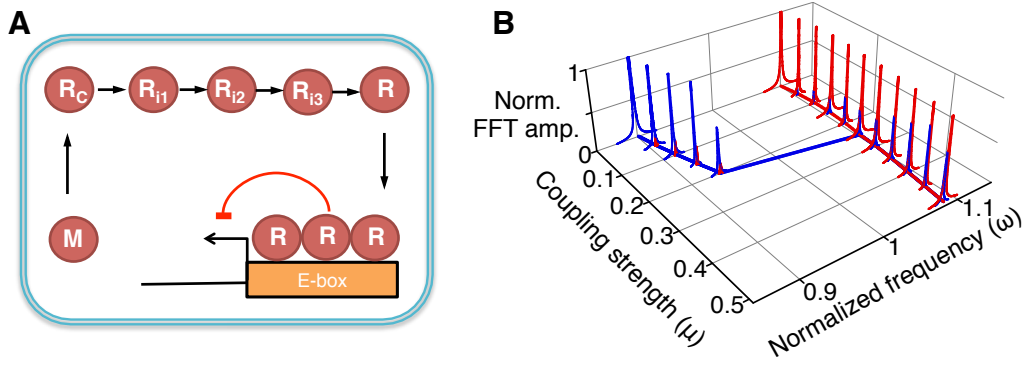

**Figure S5. The extended HT model with low Hill coefficient.** Three more intermediate steps ($R_{i1}$, $R_{i2}$, and $R_{i3}$) are included between $R_C$ and $R$ of HT model (Fig. 2B), so that the model can oscillate with a Hill coefficient of 3 (3). When a fast cell and a slow cell of the extended model are coupled, the coupled frequency is greater than the mean frequency, as in the original HT model (Fig. 4B). Parameters are the same as Fig. 4B except that Hill-coefficient is reduced to 3. The extended HT model has the form

$$\frac{dM}{dt} = \frac{1}{1+(R/K_d)^3} - M,$$

$$\frac{dR_c}{dt} = M - R_c,$$

$$\frac{dR_{i1}}{dt} = R_c - R_{i1},$$

$$\frac{dR_{i2}}{dt} = R_{i1} - R_{i2},$$

$$\frac{dR_{i3}}{dt} = R_{i2} - R_{i3},$$

$$\frac{dR}{dt} = R_{i3} - R.$$

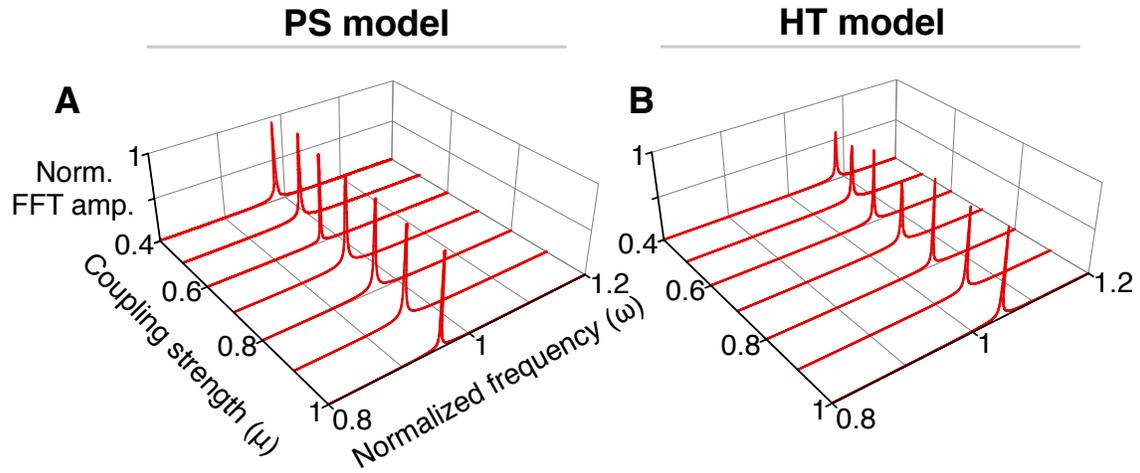

**Figure S6. The coupled periods of 100 heterogeneous cells with a stronger coupling.** Parameters are the same with Fig. 4C-D except for the coupling strengths.

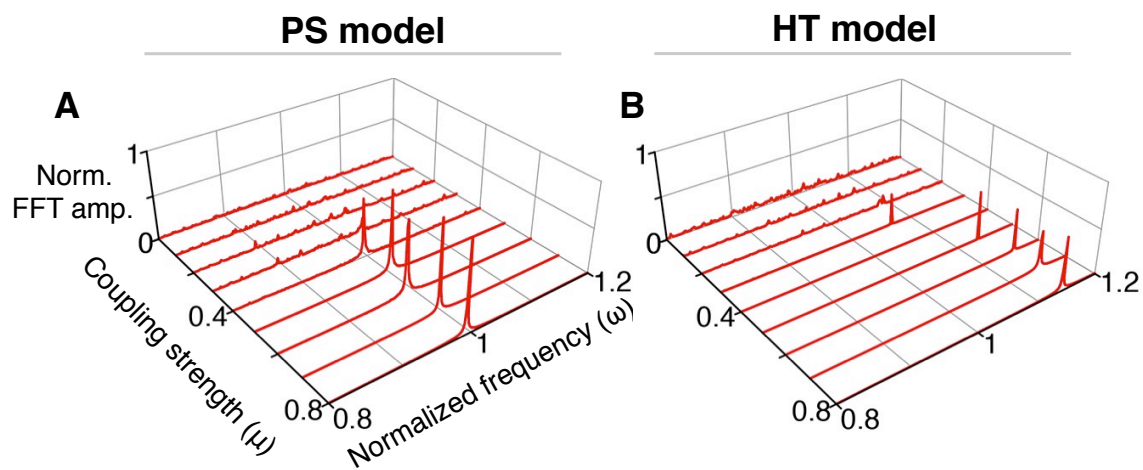

**Figure S7. The coupled periods of 100 cells with more heterogeneity.** Here, the standard deviation of rescaling factors is 0.2, which is larger than Fig. 4C-D. Other parameters are the same with Fig. 4C-D.

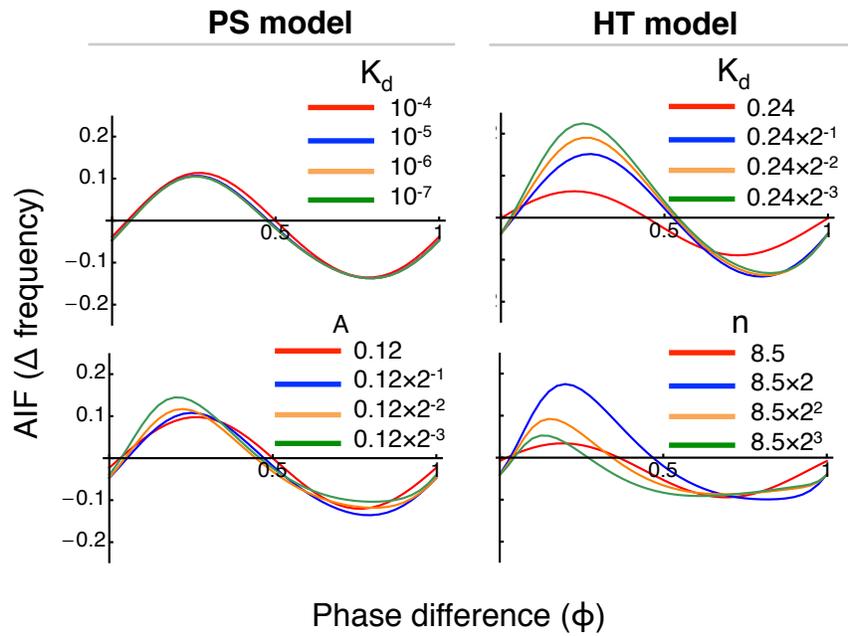

**Figure S8. The AIFs with different parameters.** Whereas the balanced AIFs of PS models are well maintained over a wide range of parameters, the AIFs of HT model show significant change. Each parameter is perturbed from the value, near which bifurcation occurs and the model begins oscillation.

## Supporting Reference